\documentclass{IOS-Book-Article}

\usepackage{mathptmx}
\usepackage{multirow}
\usepackage[center]{caption}
\usepackage{graphics}
\usepackage{url}
\usepackage{cite}
\usepackage{amsmath}
\usepackage{color}

\def\hb{\hbox to 10.7 cm{}}

\begin{document}
 
\pagestyle{headings}
\def\thepage{}

\begin{frontmatter}              

\title{Identification of Rhetorical Roles of Sentences in Indian Legal Judgments}


\author[A]{Paheli Bhattacharya\thanks{Equal contribution by the first and second authors.}\thanks{Corresponding Author: Paheli Bhattacharya; Email: paheli.cse.iitkgp@gmail.com}
}
\author[A]{Shounak Paul}$^{,1}$
\author[B]{Kripabandhu Ghosh}
\author[A]{Saptarshi Ghosh}
and
\author[C]{Adam Wyner}

\address[A]{Indian Institute of Technology Kharagpur, India}
\address[B]{Tata Research Development and Design Centre (TRDDC) Pune, India}
\address[C]{Swansea University, United Kingdom}


\begin{abstract}
Automatically understanding the rhetorical roles of sentences in a legal case judgement is an important problem to solve, since it can help in several downstream tasks like summarization of legal judgments, legal search, and so on.  
The task is challenging since legal case documents are usually not well-structured, and these rhetorical roles may be subjective (as evident from variation of opinions between legal experts). 
In this paper, we address this task for judgments from the Supreme Court of India. We label sentences in 50 documents using multiple human annotators, and perform an extensive analysis of the human-assigned labels. 
We also attempt automatic identification of the rhetorical roles of sentences. 
While prior approaches towards this task used Conditional Random Fields over manually handcrafted features, 
we explore the use of deep neural models which do not require hand-crafting of features. Experiments show that neural models perform much better in this task than baseline methods which use handcrafted features. 
\end{abstract}

\end{frontmatter}


\section{Introduction} \label{sec:intro}
\vspace{-3mm}

Rhetorical role labelling of sentences in a legal document refers to understanding what semantic function a sentence is associated with, such as facts of the case, arguments of the parties, the final judgement of the court, and so on. 
Identifying the rhetorical roles of sentences in a legal case document can help in a variety of downstream tasks like semantic search~\cite{nejadgholi2017semi}, summarization~\cite{saravanan-etal-2008-automatic, farzindar2004letsum}, case law analysis~\cite{Savelka2018SegmentingUC}, 
and so on. 
However, legal case documents are usually not well structured~\cite{DBLP:journals/ail/ShulayevaSW17, bhattacharya2019comparative}, and 
various themes often interleave with each other. 
For instance, the reason behind the judgment (Ratio of the decision) often interleaves with Precedents and Statutes. Hence it sometimes becomes difficult even for human experts to understand the intricate differences between the rhetorical roles.
Hence, {\it automating} the identification of these rhetorical roles is a challenging task.

For supervised machine learning of the roles, it is important to develop a high quality gold standard corpus, capturing the rhetorical roles of sentences as accurately as possible. 
Different approaches for the task have constructed their own set of annotated documents~\cite{nejadgholi2017semi,saravanan-etal-2008-automatic, Savelka2018SegmentingUC}, but do not report an extensive analysis on the annotation process. 
Apart from Inter-Annotator Agreement (IAA) scores, it is useful to understand issues such as the amount of subjectivity associated to the labels.
In this paper, we perform a systematic annotation study and an extensive inter-annotator study. We show that even legal experts find it difficult to distinguish some specific pairs of labels, thus showing that some subjectivity is inherent in these labels.

Prior attempts to automate the identification of rhetorical roles of sentences in legal documents~\cite{saravanan-etal-2008-automatic, farzindar2004letsum, Savelka2018SegmentingUC} rely on hand-crafted features (see Section~\ref{sec:related} for details) such as linguistic cue phrases indicative of a particular rhetorical role~\cite{saravanan-etal-2008-automatic,farzindar2004letsum, thesis}, the sequential arrangement of labels~\cite{saravanan-etal-2008-automatic}, and so on. 
Some of these features, e.g., indicator cue phrases, are {\it largely dependent on legal-expert knowledge} which is expensive to obtain. 
Also, the hand-crafted features developed in the prior works are often specific to one or a few domains/categories (e.g., 
Cyber crime and Trade secrets in~\cite{Savelka2018SegmentingUC}).
It has not been explored whether one can devise a set of features that works for documents across domains.

Recently developed deep learning, neural network models do not require hand-engineering features, but are able to automatically learn the features, given sufficient amounts of training data. 
Additionally, such models perform better in tasks like classification than methods using hand-crafted features.

In this paper, we explore two neural network models to automatically identify the rhetorical roles of sentences in legal documents --
(i)~a Hierarchical BiLSTM model, and (ii)~a  Hierarchical BiLSTM-CRF model. 
Similar models have been used in the medical domain~\cite{jin-medical-sequence-classfn}, but to our knowledge, this work is the first to use them in the legal domain.
We use these models for supervised classification across {\it seven rhetorical labels} (classes) and over documents from {\it five  different legal domains}. 
The Hierarchical BiLSTM-CRF model achieves a very good performance (Macro F-score in the range $[0.8-0.9]$), out-performing baseline methods that use hand-crafted features. 
We also analyse the rhetorical roles predicted by our model, and 
find that the subjectivity between certain pairs of labels (e.g., Ratio vs. Precedent) that is present among the human annotators is also reflected in the predictions by the algorithm. 

This is the first paper on identifying rhetorical roles of sentences in legal documents that brings together (i)~an extensive annotation study, and (ii)~deep learning models for automating the task.
\footnote{The dataset and implementations of the proposed neural model are available at {\it https://github.com/Law-AI/semantic-segmentation}.}
\section{Related Work} \label{sec:related}
\vspace{-3mm}
In this section we discuss prior work about annotation, automatic rhetorical labelling, and applications of deep learning in the legal domain.

Automatic labelling of the rhetorical role of sentences relies heavily on manual annotation. While papers that aim to automate the task of semantic labelling also perform an annotation analysis~\cite{Savelka2018SegmentingUC, DBLP:journals/ail/ShulayevaSW17}, other works focus on the process of annotation -- developing a manual/set of rules for annotation, inter-annotator studies, curation of a gold standard corpus, and so on. 
TEMIS, a corpus of 504 sentences, that were annotated both syntactically and semantically, was developed in~\cite{venturi2012design}. 
An in-depth annotation study and curation of a gold standard corpus for the task of sentence labelling can be found in~\cite{Wyner2013ACS}, where assessor agreement was low for labels like Facts and Reasoning Outcomes. 
Towards automating the annotation task, \cite{annoauto} discusses an initial methodology using NLP tools on 47 criminal cases drawn from the California Supreme Court and State Court of Appeals. 

There have been several prior attempts towards automatically identifying rhetorical roles of sentences in legal documents.
Initial experiments for understanding the rhetorical/thematic roles in court case documents/judgements/case laws were developed as a part of achieving the broader goal of summarizing these documents~\cite{farzindar2004letsum, hachey2006extractive, saravanan-etal-2008-automatic}. 
For instance, Saravanan et al.~\cite{saravanan-etal-2008-automatic} used
Conditional Random Fields (CRF)~\cite{lafferty2001conditional} for the task on 7 rhetorical roles. 
Segmenting a document into functional (Introduction, Background, Analysis and Footnotes) and issue-specific parts (Analysis and Conclusion) was looked into by~\cite{Savelka2018SegmentingUC} on U.S. court documents using CRF with handcrafted features. 
A method for identification of factual and non-factual sentences was developed in~\cite{nejadgholi2017semi} using fastText classifier. 
In another line of work, Walker et al~\cite{walker-rhetorical-roles} compared use of rule-based scripts (that require much lesser amount of training data) with Machine Learning approaches for the rhetorical role identification task.



Almost all prior attempts towards automatic identification of rhetorical roles in the legal domain have used handcrafted features. 
In contrast, this paper uses Deep Learning models for this task, where no handcrafted features are needed.
Deep Learning (DL) methods are increasingly being applied in the legal domain, e.g., classification of factual and non-factual sentences in a legal document~\cite{nejadgholi2017semi}, crime classification~\cite{wang2018modeling, wang2019hierarchical}, summarization~\cite{Liu, bhattacharya2019comparative} and other tasks.
But, to our knowledge, DL methods have not yet been applied to the task of automatically identifying rhetorical roles of sentences in legal documents.
\section{Dataset}
\label{sec:dataset}
\vspace{-3mm}

In this paper, we consider legal judgments from the Supreme Court of India, crawled from the website of Thomson Reuters Westlaw India (\url{http://www.westlawindia.com})\footnote{We use only the publicly available full text judgement. All other proprietary information had been removed before performing the experiments.}. We crawled $53,210$ documents in total. Westlaw assigns each document a legal domain, such as `Criminal', `Constitutional', etc. 
We calculated the frequency of these domains, chose the top $5$ domains and randomly sampled $50$ documents from these $5$ domains in proportion to their frequencies.
Thus we have the following set of $50$ documents from $5$ domains -- 
(i)~Criminal -- 16~documents (ii)~Land and property -- 10~documents (iii)~Constitutional-- 9~documents (iv)~Labour and Industrial -- 8~documents (v)~Intellectual Property Rights -- 7~documents. All experiments reported in this paper are performed on these $50$ case documents.
\section{Annotation Details}
\vspace{-3mm}

In this section we shall describe our annotation study, covering the rhetorical roles / semantic labels we consider in this work, the annotation procedure, and finally, analysis of inter-annotator agreement.

\subsection{Annotation Labels / Rhetorical Roles}
\label{sec:anno_labels}
\vspace{-3mm}

Our annotators were three senior Law students from the Rajiv Gandhi School of Intellectual Property Law, India ({\it http://www.iitkgp.ac.in/department/IP}).
Based on discussions with the annotators, we consider the following seven (7) rhetorical roles in our work. 

\begin{enumerate}
    \item {\bf Facts (abbreviated as FAC)}: This refers to the chronology of events that led to filing the case, and how the case evolved over time in the legal system (e.g., First Information Report at a police station, filing an appeal to the Magistrate, etc.)  
    
    \item {\bf Ruling by Lower Court (RLC)}: Since we  are considering Supreme Court case documents, there were some judgements given by the lower courts (Trial Court, High Court) based on which the present appeal was made (to the Supreme Court). The verdict of the lower Court and the ratio behind the judgement by the lower Court was annotated with this label.

    \item {\bf Argument (ARG)}: The Court's discussion on the law that is applicable to the set of proven facts by weighing the arguments of the contending parties.
    
    \item {\bf Statute (STA)}: Established laws, which can come from a mixture of sources -- Acts , Sections, Articles, Rules, Order, Notices, Notifications, Quotations directly from the bare act, and so on.

    \item {\bf Precedent (PRE)}: Prior case documents. Instructions similar to statute citations.

\item {\bf Ratio of the decision (Ratio)}:   Application of the law along with reasoning/rationale on the points argued in the case; Reason given for the application of any legal principle to the legal issue.

\item {\bf Ruling by Present Court (RPC)}: Ultimate decision / conclusion of the Court following from the natural / logical outcome of the rationale
\end{enumerate}

\subsection{Annotation Process}
\vspace{-3mm}

The annotataors used GATE Teamware tool~\cite{Cunningham2002} to annotate the documents, following the methodology of~\cite{DBLP:journals/ail/ShulayevaSW17,Wyner2013ACS}. 
An annotation manual was developed in discussion with the annotators, containing 
descriptions and example sentences for each rhetorical role, along with other instructions (e.g., a label should be assigned to a full sentence and not a part of it, a sentence should have only one label, etc.).
Initially, each annotator was asked to annotate 5 documents independently, i.e., without consulting each other. Then we had a joint discussion with all the annotators to resolve any issues, and refined the manual if necessary. 
This process was followed iteratively for annotation of the $50$ documents.

\subsection{Analysis of Inter-Annotator Agreement and Curation of Gold Standard}
\label{sec:sent_iaa}
\vspace{-3mm}

We compute the Inter-annotator Agreement (IAA) for the annotation task as follows. 

\vspace{2mm}
\noindent \underline{\textbf{IAA measure}}: As noted in~\cite{Wyner2013ACS}, aggregated pairwise Precision, Recall and F-measure are more suitable measures for IAA than measures like Kappa. 
Following the same line, we compute these pairwise IAA measures using GATE's Annotation Diff tool.\footnote{\url{https://gate.ac.uk/sale/tao/splitch10.html}} Since we have three annotators ($A_1$, $A_2$ and $A_3$), we compute three sets of pairwise IAA ($A_1$, $A_2$), ($A_2$, $A_3$), ($A_1$, $A_3$), and then take the average of the three sets. 
We briefly define the metrics below. 

GATE maintains three counts based on the extent to which two annotators' labels match. The three counts are as follows --
{\bf (1)~Correct:} If for a sentence, the two annotators mark exactly the same span of text (covering all the words and punctuation marks) with the same label, then this is considered a Correct match.  
{\bf (2)~Partial:} If for a sentence, the two annotators mark the same label, but a different span of text (e.g., leaving few words or punctuation marks), then this sentence is considered a partial match.
{\bf (3)~Missing and Spurious:} If for a sentence, the two annotators mark different labels, they are called \textit{missing} or \textit{spurious} (both terms used interchangeably). 
Based on the above definitions, Precision, Recall and F-score are calculated as follows:\\
$Precision = ( Correct + 0.5\times Partial ) / ( Correct + Spurious + Partial )
\\
Recall =  ( Correct + 0.5\times Partial ) / ( Correct + Missing + Partial )
\\
Fscore = ( ( \beta ^{2} + 1 ) \times Precision\times Recall ) / ( (\beta ^{2}\times Precision) + Recall )$
\\
where $\beta$ is the weighting of Precision vs. Recall. We use the default value of $1$, meaning that both are weighed equally. 
For each of the Precision, Recall and F-score measures, GATE computes three variants as follows --
{\bf (1)~Strict measure:} considers all partial matches as {\it incorrect} (spurious),
{\bf (2)~Lenient measure:} considers all partial matches as {\it correct}, and
{\bf (3)~Average measure:} average of the strict and lenient measures.

\begin{table}[tb]
\caption{Average inter-annotator agreement of the 3 annotators in terms of F-score as measured by GATE tool}
\label{tab:main_iaa}
\begin{tabular}{|c|c|c|c|c|c|c|c|}
\hline
\textbf{Labels $\rightarrow$} & \textbf{ARG} & \textbf{FAC} & \textbf{PRE} & \textbf{Ratio} & \textbf{RLC} & \textbf{RPC} & \textbf{STA} \\ \hline
\textbf{Strict}               & 0.692        & 0.716        & 0.654        & 0.677          & 0.74         & 0.654        & 0.857        \\ \hline
\textbf{Lenient}              & 0.953        & 0.934        & 0.878        & 0.908          & 0.925        & 0.968        & 0.967        \\ \hline
\textbf{Average}              & 0.823        & 0.817        & 0.814        & 0.821          & 0.819        & 0.798        & 0.898        \\ \hline
\end{tabular}
\end{table}

\vspace{2mm}
\noindent \underline{\textbf{Analysis of F-scores}}: 
We primarily report the F-scores, since they combine both the Precision and Recall scores.
The F-score IAA values computed by using GATE's Annotation Diff tool are presented in Table~\ref{tab:main_iaa}. 
As is expected, the strict scores are low and the lenient scores are quite high. 
This is due to differences in how different annotators use the graphical interface of the GATE tool.
For instance, one of the annotators may have mistakenly excluded the full-stop (end of sentence marker) in a sentence while marking the label, while the other annotator included the full-stop.\footnote{Though clear instructions were given to include the end of sentence marker in the label, the annotators committed this mistake while marking some of the sentences.} 
The lenient method does not take into account these errors while the strict measure does. 

Table~\ref{tab:main_iaa} reports the IAA (F-score values) for each rhetorical role individually.
In terms of strict scores, we observe Statute, Ruling by Lower Court and Facts have a high agreement whereas the scores are lower for Precedent and Ruling by Present Court. 
But in terms of lenient scores, all labels show high IAA of over $0.85$.
These IAA scores are comparable with what has been reported in similar prior studies~\cite{Wyner2013ACS}.

\vspace{2mm}
\noindent \underline{\textbf{Analysis of sentence-level agreement:}}
To understand in more detail where the annotators tend to disagree, we perform a {\it sentence-level agreement study}. 
We construct an {\it agreement matrix} $C$ (whose rows and columns are the labels) for two annotators $A_x$ and $A_y$. 
An entry $C[i][j]$ of this matrix denotes the number of sentences which Annotator $A_x$ labeled as $L_i$, but Annotator $A_y$ labeled the {\it same} sentences as label $L_j$.
Table~\ref{tab:sent_iaa_2} shows this agreement matrix for the annotator pair $(A_2, A_3)$ who have the {\it lowest IAA} (as reported by GATE).
Similar tables for the annotator pairs $(A_1, A_2)$ and $(A_1, A_3)$ are given in the Supplementary Information accompanying this paper.\footnote{Supplementary Information: \url{http://cse.iitkgp.ac.in/~saptarshi/docs/Bhattacharya-et-al-JURIX19-SuppleInfo.pdf}}




The high values in the diagonal elements indicate that the annotators have a high overall agreement in general. 
Among the non-diagonal elements, we see relatively high values (signifying some disagreement or subjectivity) for some label-pairs. 
For instance, there is subjectivity among the label pairs~(PRE, Ratio), (FAC, Ratio), (RLC, Ratio) and (RPC, Ratio), since
the reason behind the final judgement (Ratio) depends on the facts~(FAC), as well as judgements in prior cases~(PRE) and the Ruling in the lower courts~(RLC).
There is also a tendency of annotators to differ between the labels~(ARG, FAC) because framing the arguments relies on the facts of the case.

\begin{table}[tb]
\centering
\scriptsize
\caption{Table showing the the sentence level agreement between the two annotators ($A_2$, $A_3$) who have the lowest IAA (0.79, as measured by GATE)}
\label{tab:sent_iaa_2}

\begin{tabular}{|l|c|c|c|c|c|c|c|}
\hline
$\mathbf{A_2 \downarrow A_3 \rightarrow}$   & \textbf{FAC} & \textbf{ARG} & \textbf{PRE} & \textbf{STA} & \textbf{Ratio} & \textbf{RLC} & \textbf{RPC} \\ \hline
\textbf{FAC}    & \textbf{2154} & 5  & 0  & 3  & \underline{40}  & 8     & 0        \\ \hline
\textbf{ARG}    & \underline{17} & \textbf{822} & 16   & 1    & 0      & 0     & 0   \\ \hline
\textbf{PRE}    & 0   & 11    & \textbf{1425}  & 0  & \underline{47}  & 0     & 0    \\ \hline
\textbf{STA}  & 0    & 0    & 0   & \textbf{635}  & 12  & 2    & 0  \\ \hline
\textbf{Ratio}   & 4   & 13   & 4     & 5   & \textbf{3499}  & 1   & 0     \\ \hline
\textbf{RLC}   & \underline{47}  & 1   & 0   & 0   & \underline{25}  & \textbf{294}  & 0   \\ \hline
\textbf{RPC} & 6 & 0   & 0  & 0    & \underline{21}    & 0    & \textbf{262}     \\ \hline
\end{tabular}
\end{table}

\vspace{2mm}
\noindent \underline{\textbf{Analysis of agreement across domains:}}
Since we have documents from five domains of law, we 
checked the average IAA F-score for the labels across each domain. 
We found that inter-annotator agreement is uniform across different domains. Detailed results can be found in the Supplementary Information.

\vspace{2mm}
\noindent \underline{\textbf{Curation of the gold standard:}} The gold standard dataset was curated as follows: For a particular sentence, we took a {\it majority voting} of the labels given by the 3 annotators. 
There was a clear majority verdict regarding the label (rhetorical role) of each sentence. 
We use this annotated dataset in our experiments to automate the task of assigning semantic roles to sentences. 
Further statistics of the dataset are given in 
the Supplementary Information.

\vspace{-3mm}
\section{Methods for automatically identifying rhetorical roles}
\vspace{-3mm}

Now we describe our efforts towards automating the task of identifying rhetorical roles of sentences in a legal document. We treat this problem as a 7-class sequence labeling problem, where supervised Machine Learning models are used to predict one label (rhetorical role) for every sentence in a document. 

\noindent {\bf Pre-processing the documents:}
Each document was split into sentences using the SpaCy tool (\url{https://spacy.io/}). 
Splitting a legal document into sentences is challenging due to frequent presence of abbreviations~\cite{legal-sentence-boundary}. 
We observed SpaCy to do a reasonably good splitting (accuracy close to 90\%), which agrees with observations in prior works~\cite{nejadgholi2017semi}.
There were $9,380$ sentences in total in these $50$ documents, as identified by SpaCy.
Each such sentence was considered a unit for which one label (out of the seven rhetorical roles) is to be predicted.


\vspace{2mm}
\noindent \underline{\bf Baseline: CRF with handcrafted features:}
As stated in Section~\ref{sec:related}, this is the approach adopted in most prior works. 
Each document is treated as a sequence of sentences. Some dependencies exist in the corresponding sequence of labels; e.g., RLC usually follow FAC, RPC is always the end label, etc. 
Conditional Random Fields (CRFs)~\cite{lafferty2001conditional} can be used to model such sequences, since they consider both {\it emission scores} (probability of a label given the sentence) and {\it transition scores} (probability of a label given the previous label) while generating the label sequence. 

To implement the baseline approaches~\cite{saravanan-etal-2008-automatic, Savelka2018SegmentingUC}, we represent each sentence as a vector of all features stated in these works -- parts-of-speech tags (used in~\cite{Savelka2018SegmentingUC}), layout features (used in both~\cite{saravanan-etal-2008-automatic, Savelka2018SegmentingUC}), presence of cue phrases (used in~\cite{saravanan-etal-2008-automatic}), and occurrence of named entities like Supreme Court, High Court in the sentence (used in~\cite{saravanan-etal-2008-automatic}). 
The CRF works on these vectors to predict the labels (rhetorical roles).

We consider three baseline approaches: (1)~CRF using the features of~\cite{saravanan-etal-2008-automatic};
(2)~CRF using the features of~\cite{Savelka2018SegmentingUC}; and
(3)~CRF using a combination of features from both~\cite{saravanan-etal-2008-automatic, Savelka2018SegmentingUC}.

\vspace{2mm}
\noindent \underline{\bf Neural model 1: Hierarchical BiLSTM Classifier:}
We use a hierarchical BiLSTM (Bi-directional Long Short Term Memory) architecture~\cite{graves-bilstm} to automatically extract features for identifying the rhetorical roles.
This requires us to feed the sequence of sentence embeddings to the BiLSTM, which returns a sequence of feature vectors. 
The BiLSTM model needs some initialization of the sentence embeddings, with which learning can start. We try two variations of sentence embeddings: 
(1)~We construct sentence embeddings from {\it randomly initialized} word embeddings using another BiLSTM;
and, (2)~We used a large set of documents from the same domain to construct {\it pre-trained sentence embeddings}. 
Specifically, we used {\it sent2vec}~\cite{pgj2017unsup} to construct the sentence embeddings from the set of ~53K court case documents that we had collected  (see Section~\ref{sec:dataset}), {\it excluding the 50 documents considered for this task}.


\vspace{2mm}
\noindent \underline{\bf Neural model 2: Hierarchical BiLSTM CRF Classifier:}
The probability scores generated by the above model do not take into account label dependencies, and thus can be regarded as simple {\it emission scores}. To enrich the model further, we deploy a CRF on top of the Hierarchical BiLSTM architecture. 
This CRF is fed with the feature vectors generated by the top-level BiLSTM. 
As described above, we try both variations of sentence embeddings -- randomly initialized embeddings, and pre-trained embeddings trained over a large set of legal documents.
\vspace{-3mm}
\section{Results and Analysis}
\vspace{-3mm}

We now compare the performance of the models (stated in the previous section) on the set of $50$ manually-annotated documents (described in Section~\ref{sec:sent_iaa}).

\subsection{Experimental setup and evaluation metrics}
\vspace{-3mm}

We perform 5-fold cross validation with the $50$ documents, which is a standard way of evaluating Machine Learning models. 
In each fold, we have $40$ documents for training the model, and the other $10$ documents for testing the performance of the model. 
The performance measures reported are averaged over all five folds.

\vspace{2mm}
\noindent {\bf Evaluation metrics:}
For a particular sentence, the label (rhetorical role) predicted by a model is considered to be correct, if it matches with the label assigned by the majority opinion of the human annotators (see Section~\ref{sec:sent_iaa}).
We use standard metrics for evaluating the performance of algorithms -- macro-averaged Precision, Recall and F-score. 
For {\it macro-averaged metrics}, we compute these metrics for each class separately, and then take their average (to prevent any bias towards the high-frequency classes). 

\subsection{Comparing performances of different models}
\vspace{-3mm}

The comparative results are presented in Table~\ref{tab:main_results}. 
Clearly the neural models perform much better than the baselines, which shows that the latent features learnt by the neural models are better than the hand-crafted features used in prior works~\cite{ saravanan-etal-2008-automatic,Savelka2018SegmentingUC}.

\vspace{1mm}
\noindent {\bf Effect of pre-trained embeddings:} Using pretrained embeddings (learned over a large legal corpus) shows a high improvement in performance for both the neural models, as compared to using random initializations for the embeddings. 
Since deep neural models require lot of data to learn efficiently, it is especially beneficial to use pretrained embeddings learned over large domain-specific data.

\vspace{1mm}
\noindent {\bf Effect of combining CRF with neural model:} 
Hier-BiLSTM-CRF performs only a little better than Hier-BiLSTM (both with pretrained embeddings).
This is because legal documents consist of large sequences (average of 200 sentences per document), and we have few such documents; thus the CRF is unable to learn the transition scores well. Hence, there is not much additional benefit in combining CRF with the neural model.

\begin{table}[]
\caption{Macro Precision, Recall and F-score of the baseline methods and neural network-based methods. Best performances highlighted in boldface.}
\label{tab:main_results}
\resizebox{\textwidth}{!}{

\begin{tabular}{|c|c|c|c|c|c|}
\hline
\textbf{Category}                                                                                      & \textbf{Method}          & \textbf{Variations} & \textbf{Precision} & \textbf{Recall} & \textbf{F-score}     \\ \hline
\multirow{3}{*}{\begin{tabular}[c]{@{}c@{}}Baselines\\ (CRF with handcrafted\\ features)\end{tabular}} & Features from \cite{saravanan-etal-2008-automatic}                     &           -          &   0.4138                 &     0.3308            & 0.4054          \\ \cline{2-6} 
                                                                                                       & Features from \cite{Savelka2018SegmentingUC}                      & -                    & 0.4580                    & 0.4196                & 0.3250           \\ \cline{2-6} 
                                                                                                       & Features from \cite{Savelka2018SegmentingUC} and  \cite{saravanan-etal-2008-automatic}               &  -                   & 0.5070               & 0.4358                 & 0.4352          \\ \hline
\multirow{4}{*}{Neural models}                                                          & \multirow{2}{*}{Hier-BiLSTM} & Pretrained emb      &  0.8168                 &  0.7852               & 0.7968          \\ \cline{3-6}      
 & 
& Random initialization  &  0.5358                  &     0.5254           &   0.5236              \\  \cline{2-6} & \multirow{2}{*}{Hier-BiLSTM-CRF}            & Pretrained emb      &    \textbf{0.8396}                &  \textbf{0.8098}              & \textbf{0.8208} \\ \cline{3-6} 
                                                                                                       &                                           & Random initialization  &  0.6528                  &      0.5524           & 0.5784          \\ \hline
\end{tabular}
}
\end{table}

\begin{table}[tb]
\centering
\scriptsize 
\caption{F-score of the Hier-BiLSTM-CRF model, for the different labels, and for each domain of law. The last column indicates the average F-score for each domain. The last row indicates the average F-score for each of the seven labels (rhetorical roles).}
\label{tab:pred_perf}
\resizebox{\textwidth}{!}{

\begin{tabular}{|c|c|c|c|c|c|c|c|c|}
\hline
                                                                                          & \textbf{FAC}              & \textbf{ARG}           & \textbf{Ratio}              & \textbf{STA}            & \textbf{PRE}          & \textbf{RPC} & \textbf{RLC} & \textit{\textbf{\begin{tabular}[c]{@{}c@{}}Macro Average\\ (across categories)\end{tabular}}} \\ \hline
\textbf{Constitutional}                                                                   & 0.903                       & 0.659                       & 0.909                       & 0.832                       & 0.904                       & 0.857                                                                       & 0.85                                                                      & 0.845                                                                                         \\ \hline
\textbf{Labour \& Industrial Law}                                                          & 0.776                       & 0.505                       & 0.929                       & 0.423                       & 0.728                       & 0.783                                                                       & 0.681                                                                     & 0.689                                                                                         \\ \hline
\textbf{Criminal}                                                                         & 0.836                       & 0.567                       & 0.945                       & 0.689                       & 0.891                       & 0.917                                                                       & 0.865                                                                     & 0.816                                                                                         \\ \hline
\textbf{Land \& Property}                                                                  & 0.847                       & 0.624                       & 0.908                       & 0.841                       & 0.845                       & 0.98                                                                        & 0.778                                                                     & 0.832                                                                                         \\ \hline
\textbf{Intellectual Property}                                                            & 0.832                       & 0.607                       & 0.927                       & 0.824                       & 0.901                       & 0.964                                                                       & 0.886                                                                     & 0.849                                                                                         \\ \hline
\textit{\textbf{\begin{tabular}[c]{@{}c@{}}Macro Average\\ (across labels)\end{tabular}}} & \multicolumn{1}{c|}{0.8388} & \multicolumn{1}{c|}{0.5924} & \multicolumn{1}{c|}{0.9236} & \multicolumn{1}{c|}{0.7218} & \multicolumn{1}{c|}{0.8538} & \multicolumn{1}{c|}{0.9002}                                                 & \multicolumn{1}{c|}{0.812}                                                & \multicolumn{1}{c|}{--}                                                                   \\ \hline
\end{tabular}
}
\end{table}

\subsection{Detailed analysis of the best performing model (Hier-BiLSTM-CRF)}
\vspace{-3mm}

Table~\ref{tab:pred_perf} shows the F-score values of the best performing model (Hier-BiLSTM-CRF) for each of the seven labels and the five domains.

\vspace{1mm}
\noindent 
\textbf{Performance on specific labels}: From the last row of Table~\ref{tab:pred_perf}, 
we find that the model performs the best in predicting the Ratio and Ruling by Present Court (RPC). 
Ratio has the highest fraction of sentences in the corpus (38.63\%), and this large amount of training data enabled this label to be predicted well.
Ruling by the Present Court, though having less sentences (2.79\% of the dataset), always has a fixed position -- towards the end of a document. Hence this label could also be identified well. 



The model performs satisfactorily for all other labels, except `Arguments' (F-score of 0.5924). 
The `Argument' sentences get interleaved with other labels. Additionally, only 9\% of the total number of sentences in our corpus contribute to this label. Hence the neural model did not perform well in identifying these sentences.

\vspace{1mm}
\noindent 
\textbf{Performance across Domains:} 
The last column of Table~\ref{tab:pred_perf} shows how generalizable the model is across the 5 different domains. 
The model gives consistent performance (F-score in $[0.82-0.86]$) across all the domains, except for `Labour \& Industrial law'. 
This performance is consistent with the inter-annotator agreement scores, where the IAA was low for the domain `Labour \& Industrial law' (see Supplementary Information).

\vspace{1mm}
\noindent 
\textbf{Comparing inter-annotator agreement and annotator-model agreement}: 
We now compare the agreement between the human annotators (IAA), and agreement between the model and the annotators. 
We create an agreement matrix (Table~\ref{tab:confmatrix_autom}), where the rows represent the human-assigned labels (majority opinion of the annotators), and the columns represent the labels assigned by the model.  
The {\it diagonal elements} show the number of sentences for which the model-assigned label matches with the human-assigned label. 
The {\it non-diagonal} elements $C[i][j]$ shows the number of sentences where the human-assigned label $i$ does not match the model-assigned label $j$. 

\begin{table}[tb]
\centering
\scriptsize
\caption{Label agreement matrix for labels assigned by (i)~the best performing Hier-BiLSTM-CRF model, and (ii)~majority opinion of the human annotators. 
}
\label{tab:confmatrix_autom}

\begin{tabular}{|l|c|c|c|c|c|c|c|}
\hline
\textbf{Human} $\downarrow$ \textbf{Model} $\rightarrow$ & \textbf{FAC} & \textbf{ARG} & \textbf{Ratio} & \textbf{STA} & \textbf{PRE} & \textbf{RPC} & \textbf{RLC} \\ \hline
\textbf{FAC} & \textbf{1986}  & \underline{109}     & 43    & 28     & 35  & 0   & 18  \\ \hline
\textbf{ARG} & \underline{265}   & \textbf{455}     & 49   & 22      & 52  & 0   & 2    \\ \hline
\textbf{Ratio}  & \underline{129}  & 51   & \textbf{3334}  & 33   & \underline{72}  & 3  & 2        \\ \hline
\textbf{STA}   & 57    & 23   & 47  & \textbf{461}  & 55   & 0  & 3      \\ \hline
\textbf{PRE}    & 16    & 46    & 64    & 11   & \textbf{1330}  & 1  & 0    \\ \hline
\textbf{RPC} & 0  & 0  & 9   & 1    & 3  & \textbf{231}   & 18     \\ \hline
\textbf{RLC}   & 33 & 5 & 7  & 0    & 7  & 8    & \textbf{256}   \\ \hline
\end{tabular}
\end{table}

We focus on the non-diagonal elements that have relatively high values. For instance, the model seems to have frequent confusion between the labels Arguments (ARG) and Facts (FAC), and between the labels Ratio, Fact and Precedence (PRE).
Comparing Table~\ref{tab:confmatrix_autom} with the IAA agreement matrix (Table~\ref{tab:sent_iaa_2}), we find that these label-pairs are exactly the ones where the IAA values were also low, i.e., there is sufficient confusion around these label-pairs even among the human annotators.
This observation suggests that these rhetorical roles are largely subjective. 
Hence it is natural that the model will also face some difficulty in identifying these subjective rhetorical roles.



\vspace{-3mm}
\section{Conclusion and Future Work}
\vspace{-3mm}

We show that deep learning models can much better identify rhetorical roles of sentences in legal documents, compared to methods using hand-crafted features. 
We also perform an extensive annotation study, and analyse the agreement between different human annotators, as well as the agreement of the model with the annotators. 


The principal advantage of neural models is that no hand-crafting of features is needed, hence expensive legal expertise is not essential. 
However, this property also poses difficulties in understanding why exactly a sentence is more likely to be assigned to one rhetorical role than the others. 
Thus, neural models trade-off explainability/transparency with the cost of hand-crafting features. 
Deep Learning models can be used for tasks like identifying rhetorical roles of sentences, if it can be assumed that achieving good performance is more important than transparency.

In future, we plan to check how deep learning models generalize across different jurisdictions, by experimenting on legal documents of other countries.


\vspace{2mm}
\noindent {\bf Acknowledgements:} The authors thank the law students who annotated the sentences.
The research is partially supported by SERB, Government of India, through the project `NYAYA: A Legal Assistance System for Legal Experts and the Common Man in India'.
P. Bhattacharya is supported by a Fellowship from Tata Consultancy Services.
\bibliography{bibliography} 
\bibliographystyle{ieeetr}

\end{document}